\documentclass[twocolumn]{aa}
\usepackage[varg]{txfonts}
\usepackage{amsmath}
\usepackage{amsfonts}
\usepackage{graphicx,color}
\usepackage{url}
\usepackage{natbib}

\bibpunct{(}{)}{;}{a}{}{,} % to follow the A&A style 

\begin{document}

\title{A dispersion excess from pulsar wind nebulae and supernova remnants: Implications for pulsars and FRBs}
\author{S. M. Straal\inst{1,}\inst{2,}\inst{3,}\inst{4}\thanks{straal@nyu.edu} 
\and L. Connor\inst{1,}\inst{2} 
\and J. van Leeuwen\inst{2,}\inst{1}}

\institute{Anton Pannekoek Institute for Astronomy, University of Amsterdam, Science Park 904, PO Box 94249, 1090 GE Amsterdam, The Netherlands 
\and ASTRON, The Netherlands Institute for Radio Astronomy, PO Box 2, 7790 AA Dwingeloo, The Netherlands
\and NYU Abu Dhabi, PO Box 129188, Abu Dhabi, United Arab Emirates
\and Center for Astro, Particle, and Planetary Physics (CAP$^3$), NYU Abu Dhabi, PO Box 129188, Abu Dhabi, United Arab Emirates}
\titlerunning{Dispersion excess from PWNe and SNRs: Implications for pulsars and FRBs}
\authorrunning{Straal, Connor \& van Leeuwen}
\date{Accepted 13 January 2020}

\def\refbf{}

\abstract{
Young pulsars and the pulsar wind nebulae (PWNe) or supernova remnants (SNRs) that surround them are some of the most dynamic and high-powered environments in our Universe. 
With the rise of more sensitive observations, the number of pulsar-SNR and PWN associations (hereafter, SNR/PWN) 
has increased, yet we do not understand to which extent this environment influences the pulsars' impulsive radio signals.
We studied the dispersive contribution of SNRs and PWNe on Galactic pulsars, 
and considered their relevance to fast radio bursts (FRBs) such as FRB 121102.
We investigated the dispersion measure (DM) contribution of SNRs and PWNe by comparing the measured DMs of Galactic pulsars in a SNR/PWN to the DM expected only from the intervening interstellar electrons, using the NE2001 model. 
We find that a two-$\sigma$ DM contribution of SNRs and PWNe to the pulsar signal exists,
amounting to $21.1 \pm 10.6$ pc\,cm$^{-3}$. The control sample of pulsars unassociated with a SNR/PWN shows no excess. 
We model the SNR and PWN electron densities for each young pulsar in our sample and show that these indeed predict an excess of this magnitude.
By extrapolating to the kind of fast-spinning, high magnetic field, young pulsars that may power FRBs, we show their 
SNR and PWN are capable of significantly contributing to the observed DM.
}

\titlerunning{DM excess in young Galactic SNR/PWN pulsars}
\authorrunning{Straal et al.}
\maketitle

\section{Introduction}
When radio signals travel through a plasma, 
the free electrons introduce a dispersive delay, progressively slowing them down to ever lower frequencies. 
This is especially noticeable in the short radio flashes produced as coherent, broadband radio emission 
in radio pulsars \citep{1968Natur.217..709H} and fast radio bursts \citep[FRBs;][]{2007Sci...318..777L}.

By separating the dispersive delays from effects with different frequency dependence (multi-path scattering, profile evolution), \refbf{the strength of the delay can be determined.} 
Expressed as a frequency independent dispersion measure (DM), this delay directly discloses the total number of free electrons that the burst encountered. 
Combining this electron column density with its distance provides insight into the electron and baryon content along the line of sight. 

With over 2000 lines of sight, pulsars are now excellent objects to aid in the determination of such phenomena as the 3D structure of our Galaxy. 
The free electrons follow the spiral arm structure of our Galaxy and provide information of any under- or over-densities. Together with independent distance measurements, such as those drawn from parallax or \ion{H}{i} line-velocity measurements, an accurate model of the Galaxy can be obtained. 

The most widely used model for the Galactic distribution of free electrons, in the interstellar medium (ISM),  
was compiled by \citet{cordes2002,cordes2003} and is called NE2001.
It links pulsar distances and DMs. 
This smooth model is made up by a thick disk, a thin annular disk, spiral arms, and a Galactic center component. To account for further structures, clumps and voids of over-and-under-densities were added: voids for ``superbubbles" for example, and clumps for over-dense regions, such as \ion{H}{ii} regions, supernova remnants (SNRs), and O-stars.

The measurement of DM is straightforward, but pulsar distances are obtained using a variety of methods. The most common methods are parallax measurements; kinematic distances based on line velocity measurements from \ion{H}{i}, \refbf{but also, though less common, from \ion{CO}{}}; or by association. 
\cite{verbiest2012} recently reanalyzed the \ion{H}{i} kinematic distances to pulsars using the latest Galactic rotation parameters. 
After correcting for the Lutz-Kelker bias and the intrinsic pulsar luminosity distribution, a Bayesian data analysis allowed \cite{verbiest2012} to provide tighter constraints on pulsar distances.
A similar approach may be taken for FRBs. If a collection 
of redshifts is obtained and dispersion local to their host galaxy 
is understood, FRBs may act as a probe of the intergalactic medium (IGM). 

Now, for both pulsars and FRBs, the environment near the source may add to the free-electron content along the line of sight. 
Especially for young pulsars, the additional DM from pulsar wind nebulae (PWNe) or SNRs may significantly pollute the DM-distance relation that is based only on the intervening ISM and, possibly, the IGM. In our galaxy, the discovery of more SNR/PWN-pulsar associations with proper distance measurements, enables observational determinations of the DM contribution of SNRs and PWNe. 
That allows us to statistically untangle the free electron contribution near the source versus the line-of-sight contributions
for newly found young pulsars.

This could have implications for the extragalactic FRBs. It could contribute to the DM excess observed there, over the Galactic contribution (see e.g., \citealt{murase2016, piro2016}) and possibly help 
test models with significant DM contributions near the source \citep{2015ApJ...807..179P, connor-2016a}. Since the discovery of FRB 121102's repetition and its association with a star forming region \citep{tendulkar2017, kokubo2017, bassa2017}, the case that a class of FRBs can arise from young neutron stars residing in their host remnant has strengthened. 
We aim to investigate the DM contribution of the pulsars' direct environment and their SNR and PWN along with determining if this can explain the DM excess observed in FRBs.

We describe in Section \ref{sec:Method} the source selection and  data fitting. Section \ref{sec:Results} contains the results, which are interpreted and discussed in Section \ref{sec:Discussion}. In Section \ref{sec:implications}, we discuss the implications of our results for FRBs and finally, in Section \ref{sec:conclusions}, we summarize our results and conclude with our findings about the most likely contributors for excess DM in SNR/PWN pulsars.

\section{Method}
\label{sec:Method}
\subsection{Source selection}
From the ATNF catalog\footnote{\url{http://www.atnf.csiro.au/research/pulsar/psrcat/} version 1.54 \citep{manchester2005}.} we selected all radio pulsars that 1) are associated with a SNR/PWN; 2) were not used in the calibration of NE2001 and 3) have an independent, not DM-derived, distance; 
obtained from parallaxes, from \ion{H}{i} or CO line-velocity measurements, or by association. Distances by \cite{verbiest2012} were used when possible, to increase the sample homogeneity. This final set of ``associated pulsars" is detailed in Table \ref{table:sources}. 

\begin{figure}
 \includegraphics[width=0.5\textwidth]{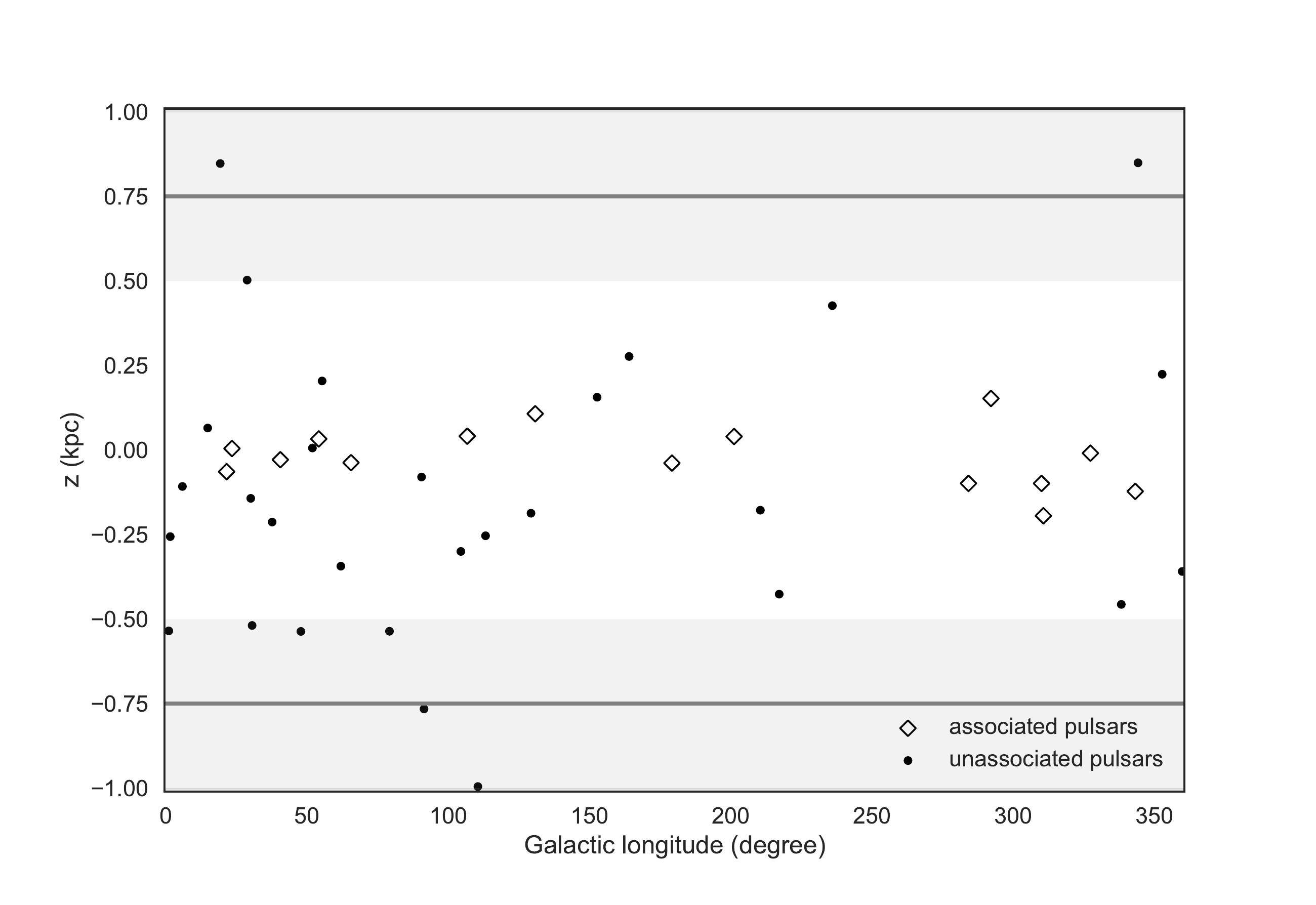}
 \caption{Height above the Galactic plane for the pulsars used in this study, without error bars. The solid line and shaded region represent one scale height of the thick disk used in the NE2001 model of $0.75\pm0.25$\,kpc.\label{fig:height}}
\end{figure}

We populated a comparison set with pulsars without SNR/PWN associations, whose distance was not included in the NE2001 calibration. 
We took care to obtain the two samples in comparable fashion; the distances were obtained using the same techniques. For our unassociated set we selected all pulsars with new parallax measurements from \citet{chatterjee2009} and \citet{matthews2016} or new \ion{H}{i} line-velocity measurements from \citet{verbiest2012}. 
To increase the sample of high-DM unassociated pulsars we include the lines of sight of pulsars in globular clusters whose scale height is less than 1.0\,kpc (see Fig.~\ref{fig:height}). 
The NE2001 model is known to have less accurate DM predictions above one modeled scale height of the thick disk of $0.75\pm0.25$ \citep{gaensler2008}. 

\begin{table*}
\caption{Pulsars associated with a SNR or PWN and with independent distance measurements. In case of multiple associations (from the ATNF catalog or other sources) the most recent work is cited.}
\label{table:sources}
\centering
\begin{tabular}{l c l l l l c}
\hline\hline
J2000 & B1950 & Distance & DM$_{\rm{obs}}$ & DM$_{\rm{NE2001}}$  & Association &  Ref.  \\%here table headers
    &    &  (kpc) & (pc cm$^{-3}$)          & (pc cm$^{-3}$) & &  \\
\hline
%\decimals
0205+6449 &   & $2.0 \pm 0.3$ & 140.7  &  $56.4\substack{+15.2 \\ -12.7}$ & PWN 3C58 & 1\\
0358+5413\tablefootmark{b}  & 0355+54 & $1.0\substack{+0.2 \\ -0.1}$ & 57.1420  &  $41.2\substack{+6.9 \\ -3.6}$ & PWN & 2 \\
0538+2817 & & $1.3 \pm 0.2$ & 39.570 & $42.4\substack{+7.3 \\ -6.8}$ & SNR S147 & 3 \\
0659+1414 & 0656+14  & $0.28 \pm 0.03$ & 14.0672 & $3.9 \pm 0.5$ & SNR Monogem Ring  & 3 \\
1016$-$5857 & & 3\tablefootmark{a} & 394.2 & $137.0\substack{+39.4 \\ -38.8}$ & PWN G284.3-1.8 & 4 \\ 
1124$-$5916 & & $5\substack{+3 \\ -2}$ & 330 & $257.5\substack{+148.3 \\ -104.9}$ & SNR G292.0+1.8 $\&$ PWN & 3 \\ 
1357$-$6429 &  & $2-2.5 \pm 0.2$ & 128.5 & $99.0(13.8) - 128.7(10.0)$\tablefootmark{c} & PWN & 5 \\
1400$-$6325 & & $6 - 8\substack{+2 \\ -1}$ & 563 & $327.2(82.8) - 426.6(58.2)$\tablefootmark{c} & SNR G310.6-1.6 $\&$ PWN & 6, 7 \\ 
1550$-$5418 & & $3.7 - 4.3 \pm 0.3$ & 830 & $183.7(41.4) - 245.3(37.6)$\tablefootmark{c} & SNR G327.24-0.13 & 8 \\
1709$-$4429 & 1706$-$44  & $2.6\substack{+0.5 \\ -0.6}$ & 75.69 & $93.35\substack{+33.3 \\ -39.8}$ & PWN G343.1-2.3 & 3 \\ 
1803$-$2137\tablefootmark{b} & 1800$-$21 &  $4.4\substack{+0.5 \\ -0.6}$ & 233.99 & $295.7\substack{+64.9 \\ -70.2}$ & SNR G8.7-0.1 $\&$ PWN &  3\\
1833$-$0827 & 1830$-$08 & $4.5\substack{+0.5 \\ -0.5}$ & 411 &  $387.9\substack{+71.2 \\ -192.9}$ & PWN & 3\\
1833$-$1034 & & 4.1$ \pm 0.3$ & 169.5 & $246.3\substack{+36.0 \\ -33.0}$ & SNR G21.5-0.9 & 3 \\
1856+0113\tablefootmark{b} & 1853+01 & 2.5$-$2.6$ \pm0.3$ & 96.74 & $65.6(16.2) - 70.8(15.4)$\tablefootmark{c} & SNR $\&$ PWN W44 & 9\\
1907+0631 & & 3.4\tablefootmark{a} & 428.6 & $91.5\substack{+39.3 \\ -34.2}$ & SNR G40.5-0.5 & 10 \\
1930+1852 & & $7\substack{+3 \\ -2}$ & 308 & $221.8\substack{141.2 \\ -83.2}$ & SNR $\&$ PWN G54.1+0.3 & 3 \\
1957+2831 & & $9.2 - 10.2 \pm 1$\tablefootmark{a} & 138.99 & $284(29.1) - 327.3(49.0)$\tablefootmark{c} & SNR G65.1+0.6 & 11 \\ 
2229+6114 & & 0.8\tablefootmark{a} & 204.97 & 10.15 $\pm 1.9$ & SNR G106.6+2.9 $\&$ PWN & 12 \\

\hline
\end{tabular}
\tablefoot{
\tablefoottext{a}{Distance error assumed to be 20$\%$ (10$\%$ on \ion{H}{i}).}
\tablefoottext{b}{For these sources a previous distance was included in the NE2001 model, but the remnant was not accounted for.}}
\tablefoottext{c}{These DMs are determined from \ion{H}{i} distance measurements and while these errors are symmetric, when translating to DM errors, they are asymmetric. Consequently, the respective errors are given for the corresponding bounds.}
\tablebib{
(1)~\citet{kothes2013}; (2)~\citet{chatterjee2004}; (3)\citet{verbiest2012}; (4)~\citet{camilo2004}; (5)~\citet{danilenko2012}; (6)~\citet{renaud2010}; (7)~\citet{marshall2006}; (8)~\citet{gelfand2007}; (9)~\citet{cox1999}; (10)~\citet{yang2006}; (11)~\citet{tian2006}; (12)~\citet{kothes2001}
}
\end{table*}

%%%% Table for unassociated pulsars
\begin{table*}
\caption{Set of pulsars unassociated with PWNe or SNRs.}
\label{table:unassociated_pulsars}
\centering
\begin{tabular}{l c l l l l c}
\hline\hline
J2000 & B1950 & Distance & DM$_{\rm{obs}}$ & DM$_{\rm{NE2001}}$  &  Association & Ref.  \\%here table headers
    &     & (kpc) & (pc cm$^{-3}$) &  (pc cm$^{-3}$) & &\\
\hline
%\decimals

0030+0451 &  & $0.30\substack{+0.02 \\ -0.01}$ & 4.33252 &  $3.72\substack{+0.67 \\ -0.34}$ &  & 2 \\
0034$-$0721 & 0031$-$07  &  $1.03\substack{+0.08 \\ -0.08}$ & 10.922 & $24.86\substack{+1.1 \\ -1.2}$ & & 1 \\
0139+5814 & 0136+57 & $2.6\substack{+0.3 \\ -0.2}$ & 73.811 & $62.29\substack{+12.3 \\ -9.7}$ & & 1\\
0452$-$1759 & 0450$-$18 & $0.4\substack{+0.2 \\ -0.1}$ & 39.903 & $5.85\substack{+4.3 \\ -1.6}$ & & 1 \\
0454+5543 & 0450+55 & $1.18\substack{+0.07 \\ -0.05}$ & 14.590 & $32.06\substack{+2.3 \\ -1.6}$ & & 1 \\
0613$-$0200 & & $1.1\substack{+0.2 \\ -0.2}$ & 38.77919 & $0.3\substack{+6.4 \\ -6.4}$ & & 2 \\
0645+5158 &  & $0.8\substack{+0.3 \\ -0.2}$ & 18.247536 & $21.4\substack{+8.7 \\ -6.5}$ & & 2 \\
0820$-$1350 & 0818$-$13  & $1.9\substack{+0.1 \\ -0.1}$ & 40.938 &$38.56\substack{+2.6 \\ -2.6}$ & & 1 \\
1600$-$3053 & & $3.0\substack{+1.0 \\ -0.6}$ & 52.3245 & $89.4\substack{+17.0 \\ -13.8}$ & & 2 \\
1614$-$2230 &  & $0.65\substack{+0.05 \\ -0.04}$ & 34.3864 & $10.8\substack{+1.9 \\ -1.5}$ & & 2\\
1713+0747 &  & $1.18\substack{+0.04 \\ -0.04}$ & 15.9780 & $24.8\substack{+1.2 \\ -1.2}$ & & 2\\
1740$-$5340	&  & $2.2\substack{+0.5 \\ -0.7}$ & 71.8 & $86.79\substack{+15.7 \\ -28.4}$ & NGC6397 & 3 \\
1744$-$1134 &  & $0.41\substack{+0.02 \\ -0.02}$ & 3.13695 & $2.96\substack{+0.77 \\ -0.05}$ & & 2\\
1801$-$0857ABCD   & & $7.2$\tablefootmark{a} & 180.48\tablefootmark{b} & $247.43\substack{+29.3 \\ -37.0} $ & NGC6517 & 4 \\
1803$-$3002ABC  & & $7.8$\tablefootmark{a} & 193.03\tablefootmark{b} & $324.01\substack{+50.8 \\ -55.9} $ & NGC6522 & 5 \\
1807$-$2459AB  & & $2.8$\tablefootmark{a} & 135.58\tablefootmark{b} & $128.36\substack{+35.7 \\ -34.1} $ & NGC6544 & 6 \\
1835$-$3259A  & & $10.7$\tablefootmark{a} & 63.35 & $198.81\substack{+3.3 \\ -7.9} $ & NGC6652 & 7\\
1909$-$3744 & & $1.07\substack{+0.04 \\ -0.03}$ & 10.3932 & $33.9\substack{+1.5 \\ -1.2}$ & & 2\\
1909+0254 & 1907+02 & $4.5\substack{+2.2 \\ -0.9}$ &  171.734 & $150.0\substack{+103.2 \\ -43.6}$ & & 1\\
1918$-$0642 &  & $0.9\substack{+0.2 \\ -0.1}$ & 26.554 & $13.1\substack{+8.0 \\ -3.9}$ & & 2\\
1922+2100 & 1920+21 & $4\substack{+2 \\ -2}$ & 217.086 &  $91.2\substack{+71.9 \\ -66.8}$ & & 1 \\
1926+1648 & 1924+16 & $6\substack{+3 \\ -2}$ & 176.885 &  $184.7\substack{+128.1 \\ -89.2}$ & & 1 \\
2043+1711 && $1.3\substack{+0.4 \\ -0.3}$ & 20.70987 & $14.9\substack{+4.8 \\ -3.6}$ & & 2\\
2048$-$1616 & 2045$-$16  & $0.95\substack{+0.02 \\ -0.03}$ & 11.456 & $23.24\substack{+0.6  \\ -0.8} $ & & 1 \\
2055+3630  & 2053+36 & $5.0\substack{+0.8 \\ -0.6}$ & 97.4155 & $122.74\substack{+29.2 \\ -23.1} $ & & 1\\
2145$-$0750 & & $0.8\substack{+0.2 \\ -0.1}$ & 8.99761 & $15.6\substack{+4.9 \\ -2.7}$ & & 2\\
2157+4017 & 2154+40  & $2.9\substack{+0.5 \\ -0.4}$ & 71.1239 & $55.06\substack{+18.8 \\ -14.6}$ & & 1 \\
2313+4253 & 2310+42  & $1.06\substack{+0.08 \\ -0.06}$ & 17.277 & $12.31\substack{+1.7 \\ -0.7}$ & & 1\\
2317+1439 &  & $1.3\substack{+0.4 \\ -0.2}$ & 21.8999 & $31.7\substack{+5.4 \\ -3.6}$ & & 2\\

\hline

\end{tabular}
\tablefoot{
\tablefoottext{a}{Distance error assumed to be 20$\%$.}
\tablefoottext{b}{The average DM is taken for these Globular Cluster pulsars.}
}
\tablebib{
(1)~\citet{verbiest2012}; (2) \citet{matthews2016}; (3) \citet{heyl2012}; (4) \citet{kavelaars1995}; (5) \citet{harris1996}; (6) \citet{valenti2010}; (7) \citet{chaboyer2000}
}
\end{table*}

\subsection{Fitting}
\begin{figure}
 \includegraphics[width=0.5\textwidth]{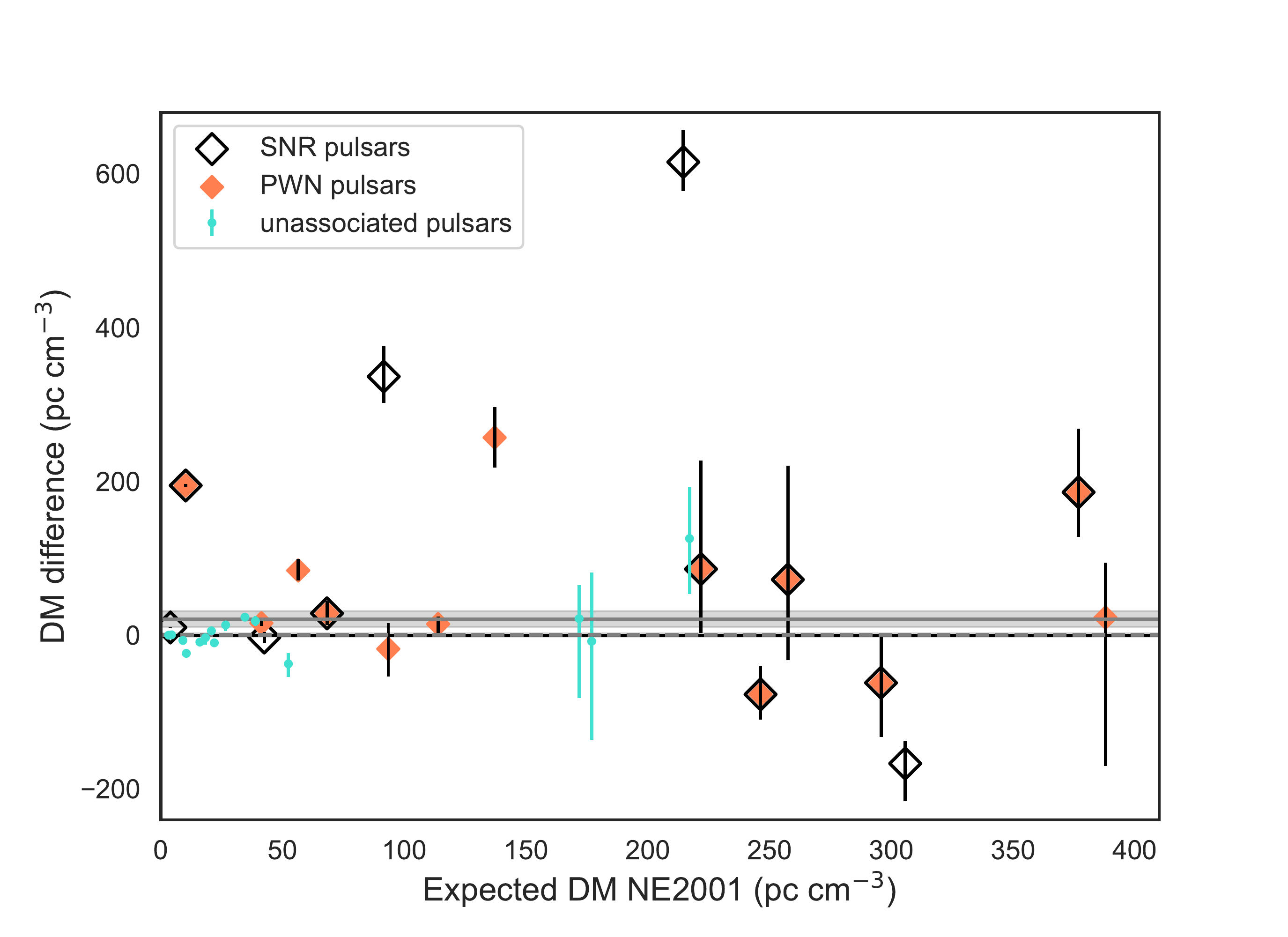}
 \caption{DM difference between measured and predicted, as a function of the NE2001 predicted DM. Associated pulsars are given in diamonds, where open, black diamonds are SNR associations, filled orange diamonds represent PWN associations and black lined diamonds with filled centres are plerionic systems. The pulsars unassociated with a SNR or PWN are shown in turquoise.
 %. Pulsars in SNRs are represented by the empty diamonds, in PWN by filled diamonds, potentially inside the SNR. 
 The offset DM for the associated sample is represented by the grey, solid line with its one $\sigma$ error-region shaded. The fit, and the one $\sigma$ error-region, to the offset for the unassociated sample is shown as the grey dashed line and overlaps with the black solid x-axis.\label{fig:results}}
\end{figure}

We take the NE2001-predicted DM at each 
source's distance to be the model DM. 
%As model DMs we define the NE2001 predictions for the pulsar distance. 
We propagated the distance errors to model DM errors, and do not include systematic errors from the NE2001 model itself, opting instead to compare our associated pulsars with a control sample.
The resulting values are listed in the distance Tables \ref{table:sources} and \ref{table:unassociated_pulsars}, described above. 
In Fig. \ref{fig:results} we show, for both our sets of pulsars, the difference between the the observed and expected DM, versus the expected DM.

Under our hypothesis that the local surroundings of the associated pulsars add to the DM, we expect an increase from the predicted DM that is unrelated to DM magnitude. Hence, we fit for such an offset, as  $\rm{DM_{obs} = DM_{exp} + b}$, where $\rm{DM_{obs}}$ and $\rm{DM_{exp}}$ are the observed and expected DM (see Tables \ref{table:sources}, \ref{table:unassociated_pulsars}), and b the offset.

We fit the data using the least-squares-method,
assuming the errors follow a Gaussian distribution. 
To be compatible with this approach, we symmetrized our error bars by (1) averaging the positive and negative error, or (2) centering the data point. Both approaches yield the same result. Distance uncertainties from \ion{H}{i} line-velocity measurements are not Gaussian, but only reflect distance upper and lower limit. We treat these as Gaussian and centered. 
To validate this approach, we also used Markov chain Monte Carlo \citep[MCMC, implemented in python as emcee,][]{emcee} fitting with a more accurate error distribution for each data point (asymmetric Gaussians for the asymmetric errors, and top hats with Gaussian drop-offs for the \ion{H}{I} obtained distances). 
We retrieve the same DM offset as with our direct least-squares-method, but the reported errors do not resemble the data variance. Therefore, we hereafter quote the results from the least-squares method.

\section{Results}
\label{sec:Results}
\begin{figure}
 \includegraphics[width=0.5\textwidth]{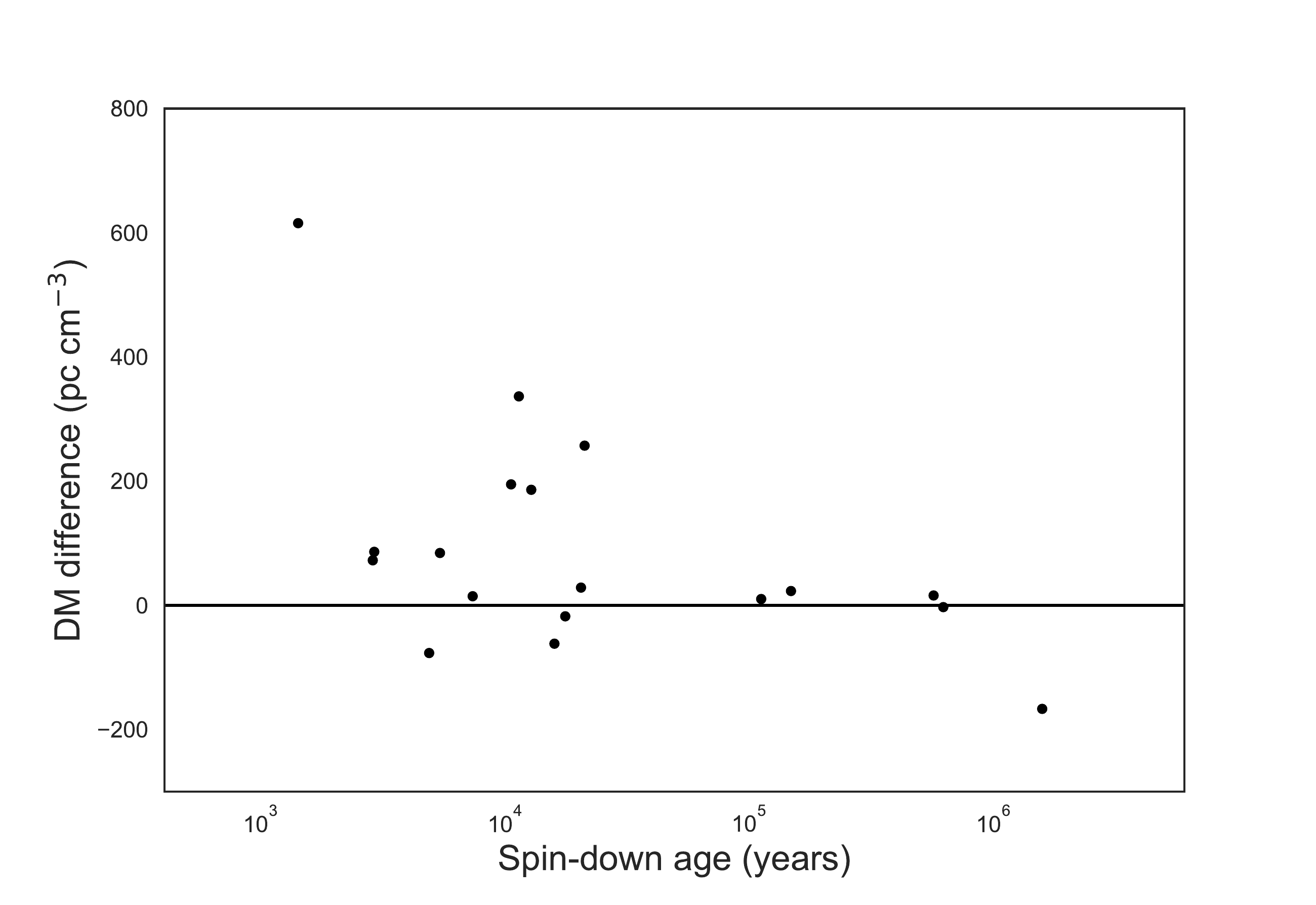}
  \caption{Spin-down age of the associated pulsar sample (black dots) shown against their DM excess (DM$_{\rm{obs}}-$DM$_{\rm{NE2001}}$).
 \label{fig:age}}
\end{figure}

Based on our a-priori expectations that PWNe and SNRs are a source of additional electrons in the line of sight, we set out to investigate if a measurable DM excess exists, and of what magnitude.
For the pulsars in a SNR/PWN we find an offset of $21.1 \pm 10.6$ pc\,cm$^{-3}$, \refbf{where the quoted uncertainty denotes the one sigma error on the given mean offset}. We check the robustness of the fit by taking out the pulsar with the highest DM-offset (PSR J1550-5418, see Table \ref{table:sources}) and find that the offset remains the same and is, hence, not dominated by this outlier.
Given this 2-sigma detection we are 95\% confident the offset is real. From here on, we call this 2-sigma offset the ``excess.'' 
In our comparison sample, no SNRs or PWNe are visible. As a result, we do not expect an excess there.
We find the comparison sample does indeed agree with the model-predicted DM: it shows no excess when determined by the 
\refbf{least-squares fitting method ($0.68 \pm 1.9$\,pc\,cm$^{-3}$) 
and even a slight deficit when determined by the 
MCMC approach ($-5.3 \pm 2.0$\,pc\,cm$^{-3}$).}
This agrees with our hypothesis that it is SNRs and PWNe, absent here, that increase  observed DMs. 
These results are shown in Fig. \ref{fig:results}. 
We checked for an age-dependence of the 
DM excess and find a weak trend (see Fig. \ref{fig:age}). 
The DM excess seems to lessen as a function of age, 
as expected due to expansion of the SNR and PWN over time.

\begin{figure}
 \includegraphics[width=0.5\textwidth]{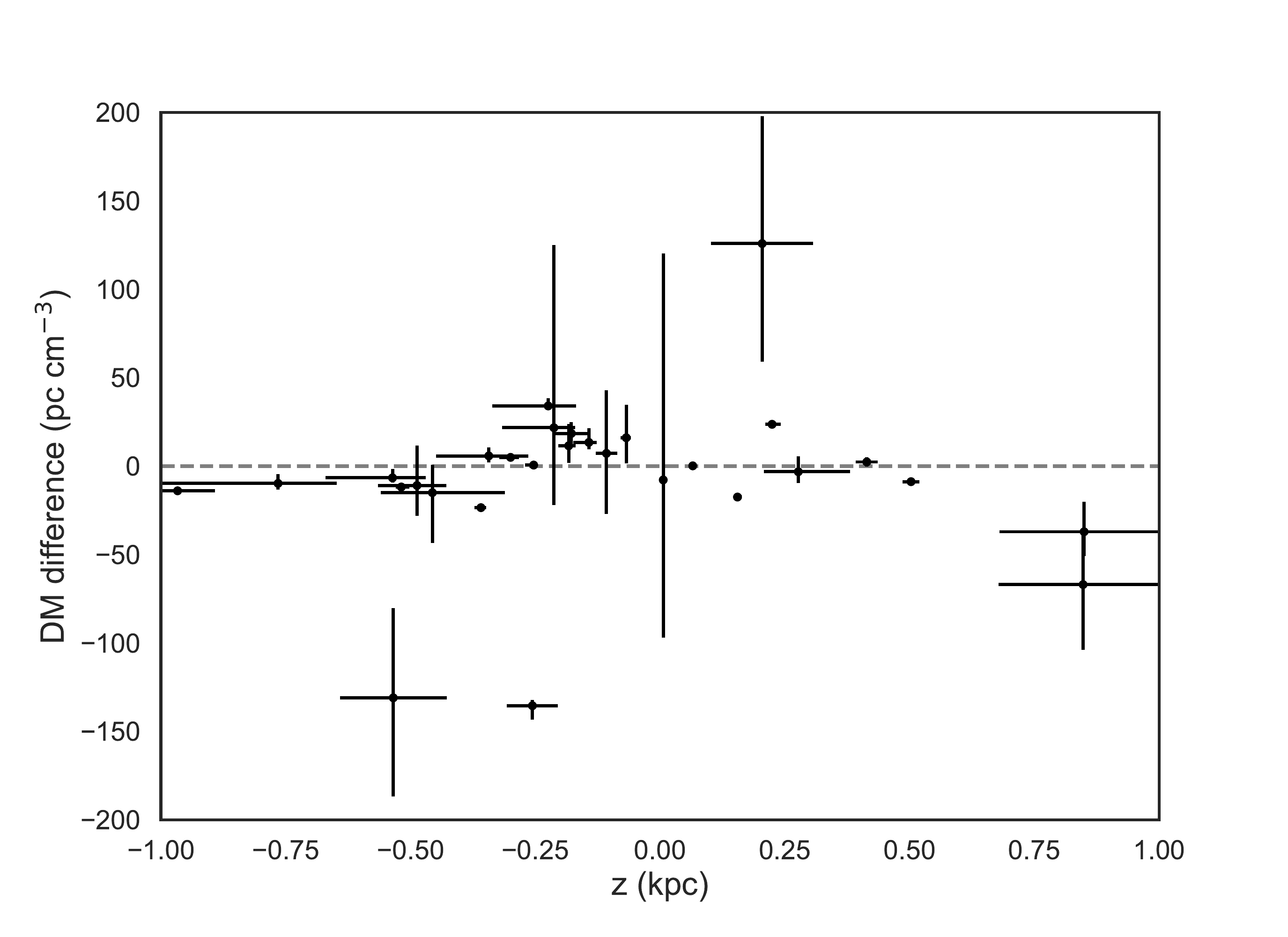}
  \caption{DM difference of pulsars unassociated with a PWN or SNR (see Table \ref{table:unassociated_pulsars}) shown against their scale height. The dashed line indicates zero DM difference.
 \label{fig:DM_scaleheight}}
\end{figure}

To validate our selection criterion on the maximum scale height for pulsars in the unassociated sample, we checked for a correlation of DM excess with height of the pulsars above the Galactic plane \refbf{by calculating the Pearson correlation coefficient.} 
\refbf{
The DM excess for the unassociated sample as a function of scale height is given in Fig. \ref{fig:DM_scaleheight}.
We determine the correlation coefficient of the DM difference and the absolute scale height to be $-0.36$.
Hence, the DM excess of the unassociated sample is not strongly correlated with scale height and thus we rule out that the on average larger scale height for the pulsars unassociated with a PWN or SNR affects our results.}
Additionally, we investigated the influence of using the Lutz-Kelker corrected distances from \citet{verbiest2012} on our results by using the uncorrected distances for the predicted DM values.
Using these uncorrected distances, our excess increases slightly to $21.3$ pc\,cm$^{-3}$. 
Given this small effect we choose to use the distances as given by \citet{verbiest2012} to keep the distance determinations of the samples as homogeneous as possible.

\section{Discussion}
\label{sec:Discussion}
We set out to investigate excess DM
in young Galactic pulsars. Beyond adding to our understanding of young pulsars, 
this could inform about the local plasma environment around FRBs; half of 
FRB 121102's extragalactic dispersion 
seems to come from within its host galaxy \citep{chatterjee2017,tendulkar2017}.
We have found such a 2-sigma excess.
Here we discuss the observational and theoretical constraints on the origin of this observed DM excess. 
We consider the following three classes for the origin of ``local'' DM.
The excess could arise from the innermost environment, in a PWN.
It could also come from a SNR, depending on its age, expansion rate, and environment.
Finally, free electrons in the pulsar surroundings, other than the nebula, could cause it: the sources in our sample are overall young, energetic pulsars and could be associated with active star-forming regions or have previously-ionized wind bubbles. Star-forming regions are known sources of \ion{H}{ii} regions. 
While these three could in principle be derived for each of the sources in our sample, we found there to be too many individually unknown quantities. For the current discussion we thus take an ensemble approach, to investigate the overall, average excess.

\subsection{Pulsar wind nebula}
\label{sec:pwn}
If the excess arises from the PWN, estimating dispersion caused by the wind must account for relativistic effects and the PWN's density structure.
\refbf{
In the pulsar magnetosphere, electron-positron pairs are created \citep{goldreich1969} which are launched at the light cylinder radius ($r_{\rm{LC}} = c / \Omega$), forming the pulsar wind.
The wind at this point is considered to be isotropic, it may form a toroidal structure or become collimated outside the wind termination shock radius, $r_{\rm{TS}}$ (see \citet{gaenslerslane2006} and references therein).
To determine whether the free electrons in the wind can contribute significantly to the observed DM, we note that the hot, shocked, relativistic plasma is not expected to contribute, due to Lorentz suppression.
However, the cold component of the wind, near the light cylinder, moves outwards with a relativistic bulk flow and can increase the DM.
Thus the contribution of the wind generated from the light cylinder up to the termination shock is evaluated.}

\refbf{Although the electron column density in the wind is Lorentz invariant \citep[$n'$d$l'=n$d$l$,][]{yu2014}, the plasma frequency is not.
In the observer's frame the plasma frequency is Doppler-boosted \citep{yu2014, cao2017} and this needs to be taken into consideration when the determining the DM contributed by the wind.}
The particle number density in this wind as a function of radius can be derived from the Goldreich-Julian density, $n_{\rm GJ}$, and a multiplicity factor, $\kappa$, which is the average number of pairs produced per streaming particle. This is given by,

\begin{equation}
n_{W}(r) = \kappa\, n_{\rm GJ}\, \left ( \frac{r}{r_{\mathrm{LC}}} \right)^{-2},
\label{eq-nwind}
\end{equation}

\refbf{
\noindent where $r_{\mathrm{LC}}$ is the light cylinder radius and for $r>r_{\rm{LC}}$. 
\cite{yang2017} investigate the dispersion in a magnetized plasma, such as the pulsar wind, including the contribution from the four independent modes of wave propagation.
They find that ``ordinary wave'', or O-mode, dominates the DM and the effective dispersion measure of this wind can be calculated as,
}

\begin{equation}
\mathrm{DM}_W = \int 2\,\Gamma (r)\, n_{W}(r)\,\mathrm{dr},
\label{eq:dm_w}
\end{equation}

\refbf{
\noindent where $\Gamma(r)$ is the Lorentz factor as a function of radius (see Eq.~8 in \citealt{cao2017}). The Lorentz factor, $\Gamma$, evolves in the wind from $\sim 10^2$ at the light cylinder radius to $\sim 10^6$ at the termination shock \citep[see][and references therein]{gaenslerslane2006}.
As the mechanism for this transition is unclear \citep{melatos1998,arons2002} and the contribution to the DM is mainly from the inner region where the number density of the electrons is higher, $\Gamma$ is evaluated at $r_{\rm{LC}}$. 
The particle number density and the Lorentz factor evaluated at $r_{\rm{LC}}$ are determined only by the neutron star's magnetic field, spin period, and the multiplicity factor $\kappa$.
Hence evaluating Eq. \ref{eq:dm_w} from $r_{\rm{LC}}$ to $r_{\rm{TS}}$ for $r_{\rm{TS}} \gg r_{\rm{LC}}$, \citet{yang2017} show that DM$_{W}$ from the cold relativistic bulk flow can be expressed as follows (here adapted to reflect a typical magnetic field for young Galactic pulsars):
}

\begin{equation}
\mathrm{DM}_W = 18 \times \left ( \frac{B}{10^{13}\,\mathrm{G}} \right )^{4/3}
\left ( \frac{P}{100\,\mathrm{ms}} \right )^{-11/3}
\left ( \frac{\kappa}{10^4} \right )^{2/3}
\,\mathrm{pc}\,\mathrm{cm}^{-3},
\label{eq-pwn}
\end{equation}

\refbf{
\noindent where $B$ is the neutron star's magnetic field and $P$ the pulsar period. 
Considering that for Galactic PWNe and pulsars, $r_{\rm{TS}}$ is typically 0.1\,pc \citep{gaenslerslane2006} and $r_{\rm{LC}}$ is of order $10^9$\,m we can apply this equation to our sample.
}
The extra DM from Eq.~\ref{eq-pwn} is strongly dependent on spin period and magnetic field. 
We find that for PWN-associated pulsars in our set, DM$_W$ ranges from $0.03-69$\,pc\,cm$^{-3}$, with a mean of 9\,pc\,cm$^{-3}$. The maximum DM$_W$ for the unassociated pulsars, as calculated from Eq.~\ref{eq-pwn}, was 4\,pc\,cm$^{-3}$. 

\refbf{
We assumed the wind to be isotropic and any deviations from that may cause the given contribution to decrease or increase depending on the orientation of the pulsar beam with respect to the nebula, and our viewing angle.
Since this can go either way, we only evaluated the wind contribution for the simplified isotropic wind.
Next, the multiplicity factor is expected to be in the order of $10^3-10^5$. 
We assumed the multiplication factor to be $\kappa=10^4$. 
Nonetheless, \citet{bucciantini2011} provide inferred lower limits on the multiplicity factor for a set of six PWNe of $>10^5$. This would increase the DM contribution of the wind by a factor of $\sim 5$. 
However, given that $\kappa$ may vary over the sample we have chosen for $\kappa=10^4$, which is representative for the sample as a whole.
Taking all into consideration, the Doppler-boosted plasma frequency of the wind could explain the trend we see towards DM-excess in associated pulsars.
} 
\citet{yang2017} do a similar calculation, but account for the fact that in a strong magnetic field, 
the electron cyclotron radiation grows large and the propagating wave must be decomposed into orthogonal modes. They also conclude that the PWN DM can become significant, particularly for a rapidly-rotating magnetar, like the suggested progenitor to FRB 121102 \citep{metzger2017}.

PWNe can also have filamentary structures: either formed by Rayleigh-Taylor instabilities in the nebula \citep{blondin2017}; or they can be inbedded in a filamentary structure from interaction with the pre-explosion circumstellar medium.
A well-known example of the latter is the Crab nebula, where the filamentary structure is believed to come from the progenitor star \citep{fesen1982}.
The over-densities in these filaments can be significant, and contain several solar masses of ionized and neutral material \citep[$4.7\pm1.8 M_{\odot}$][]{fesen1997}.
In \citet{blondin2017}, the filamentary structure in a PWN is modelled by allowing the expansion to have Rayleigh-Taylor instabilities. 
If the line of sight to the pulsar happens to be through one of these filaments, the DM can be enhanced significantly. This chance alignment does not happen in all cases. \cite{blondin2017} argue the enhancement only occurs for less than a few percent of the lines of sight and therefore it does not explain the excess for the full sample.

\subsection{Supernova remnant}
Further out from the pulsar than the PWN, supernova remnants are known to provide free electrons. Even for the sources where no SNR is observed, the environment can still contain excess ionized material from the supernova explosion. We determine the number density in the SNR using:
 \begin{equation}
 n_{\mathrm{e}} = f_i\,\frac{M_{\mathrm{ej}}}{\mathrm{4\pi}\Delta r_{\rm sh}
 r^{\mathrm{2}}\mu_{\mathrm{e}}m_{\mathrm{p}}}\,\mathrm{cm^{-3}},
 \end{equation}

\noindent where $M_{\rm{ej}}$ is the mass ejected in the supernova explosion confined 
to a shocked shell of width $\Delta r_{\rm sh}$ at radius $r$. During the Sedov-Taylor phase,

\begin{equation}
r_{\rm{s}}=\!\rm{12.9\,pc}\, (t/\rm{10^4}\,\rm{yr})^{2/5}\left(\epsilon_0/n_0\right)^{1/5},
\end{equation}

\noindent is the shock-radius of the remnant \citep{cox1972}, with $\epsilon_0 = 0.75 \times 10^{51} $ ergs as typical explosion energy and n$_0$ the ambient interstellar medium density. The mean molecular weight is given by $\mu_{\rm{e}} = 1.3$, $m_{\rm{p}}$ the proton mass, and $f_i$ the ionization fraction.
For the maximum contribution of fully ionized $8\,M_{\odot}$ ejecta mass and an equal amount of swept-up mass (total of $\sim16\,M_{\odot}$ ionized material in the shell), $n_0=1$\,cm$^{-3}$, $\Delta r_{\rm sh} = 0.05$\,pc, the extra DM at the median characteristic age of our sample of t = $10^{4}$ years, is 4.8\,pc\,cm$^{-3}$.
%and decreases to 0.06\,pc\,cm$^{-3}$ after t = $10^{5}$ years.
% While this optimisitic scenario cannot fully explain the excess, SNRs could add a few DM units to associated Galactic pulsars; for FRBs, which may be $\sim$10$^{2}$ yr (rather than the $\sim$10$^{3-5}$ yr of our sample) the remnants would be
% considerably more compact. 

We used a toy model for the ionization 
and expansion of the SNR shell, ignoring 
for now the effect of the surrounding ambient 
medium. However, in an analysis of the DM evolution of
supernova remnants by \citet{piro2018}, it is shown that 
in free expansion the DM decreases more slowly than expected, 
as $\propto t^{-1/2}$ and 
during the Sedov-Taylor phase, material swept up 
by the ISM can contribute to local 
DM such that it increases with time, 
such that DM$\propto t^{2/5}$. 
They find local DMs of $10^{-1}-10^2$\,pc\,cm$^{-3}$
for a 10,000 year old pulsar, depending on 
the density of the surrounding ISM.

% If the true SNR radius evolution is $\sim20\%$ slower than here modeled, $\sim4$\,pc\,cm$^{-3}$ of extra DM is possible.
 
\subsection{Star-formation region}
\label{sec:sfr}
Excess DM could also arise from the pulsar environment outside its PWN/SNR. The young, energetic pulsars in our sample could be associated with star-forming regions (SFRs) or reside in previously-ionized wind bubbles. However, of the 18 pulsars in our sample only two are associated with a star-forming or \ion{H}{ii} region. PSR J1550$-$5418 has two associated massive stars, and PSR J1856+0113 in SFR W48.

If the \ion{H}{ii} region is formed by massive stars, as is the case with PSR J1550$-$5418, the size of the ionized region can be determined by assuming a typical Str\"omgren sphere \citep{stromgren1939}:
\begin{equation}
R_{s} = \left(\frac{3N^*}{4\pi\alpha n_{e}^2}\right) ^{1/3}.
\end{equation} 

\noindent For an O-type star with $T_e = 40,000$\,K, the Str\"omgren radius, $R_s$, is $\sim$15\,pc, while assuming the following parameters. The density of hydrogen-ionizing photons $N^*$ is $\sim10^{49}$ photons $\rm{s^{-1}}$ \citep{sternberg2003}, the recombination rate density is given by $\alpha n^2$, with $\alpha = 2.6\times10^{-13} \rm{\,cm^3\,s^{-1}}$ the recombination coefficient for an O-type star with $T_e = 40,000$\,K \citep{spitzer1978} and the ambient electron density equals the ambient hydrogen density, $n_{\rm{H}} = n_e \sim100\, \rm{cm^{-3}}$ \citep[see][chap.~15]{stahler2005}.
With these values, $\sim$300 DM units 
can be added if a pulse 
travels through the Str\"omgren sphere. For PSR J1550$-$5418, whose excess is ~600 units, 
such bubbles could contribute significantly to its DM. The pulsar, as well as its surrounding region, are also known to have a large negative 
rotation measure (RM), which we discuss further in Sect.~\ref{sec:conclusions}.

\subsection{Other indications for excess electron density} 
The DM-scattering relation for Galactic pulsars in \cite{cordes2016} shows outliers. Interestingly, two are PWN-pulsars that are severely under-scattered, or, over-dispersed. 
The authors argue that these signals traveled through overdense regions that enhanced the DM but not the scattering, such as the PWN. The implied excess in these PWNe is 45\,pc\,cm$^{-3}$ for J1709$-$4428 and 130\,pc\,cm$^{-3}$ for J0908$-$4913. This respectively represents 60$\%$ and 73$\%$ of their observed DMs. \cite{cordes2016} also offer an alternative explanation where the overdense region is close but unaffected by the pulsar, and geometrically disfavours scattering enhancement. That may of course be occasionally valid for certain sources. But given the general trend in PWN pulsars that we presented earlier, we argue that  this enhancement is contributed by the PWN itself. 

\subsection{Limitations from the electron-density model}
The NE2001 DM predictions are known to sometimes be off by 20$-$30\%, especially in certain directions of the Galaxy that are less well modeled. In our sample, pulsars are well distributed over our Galaxy, and NE2001's predictions for the unassociated sources agree with their measured DMs. We observe no trend in Galactic latitude for DM excess and conclude these direction-dependent model variations are not the cause of our observed excess.
We show that for the unassociated pulsars there is no evidence for DM excess.
However, unassociated sources 
are on average more nearby and more likely to have parallax-determined distances; if DMs 
are systematically underestimated by NE2001 at greater distances, then errors could be introduced in our excess fit. 

Recently, a new model for the Galactic distribution of free electrons was introduced: YMW16 \citep{ymw16}. While this model shares many  similarities with the NE2001 model, one of the main differences is that it does not rely on inclusion of voids or clumps to better fit the data. In part due to the increased data sample, \cite{ymw16} show that YMW16 more accurately predicts the pulsar DM and distance than NE2001. We checked whether we could compare the YMW16 model to NE2001 for our sample. However, most of the pulsars in our sample are used in its calibration, barring us from doing an independent comparison.
As the YMW16 code and default parameter sets can be publicly downloaded we have investigated if we could derive a parameter set based on a population that excludes the associated pulsars; and then investigate the prediction of this unassociated model for the associated pulsars. However, given the codebase, we concluded that (re)implementing the pulsar input data sets, plus the fitting and verification framework around the available model code are beyond the scope of this paper.

\section{Implications for FRBs}
\label{sec:implications}
Synthesizing these above aspects, the doppler-boosted PWN ejecta combined with the ionized ejecta in SNRs offer the best explanation for the observed DM excess of $21.1 \pm 10.6$ pc\,cm$^{-3}$ in radio pulsars associated with such hosts.
However, if FRBs are produced by young neutron stars, they could also be associated with star forming regions.
Hence, we further explore the possibility of similar mechanisms explaining the excess DM observed in FRBs, albeit likely in more extreme environments.

So far, three FRBs have been localized, the repeating FRB~121102 \citep{tendulkar2017}, and the one-off FRBs~180924, and 190523 \citep[][]{bannister2019,ravi2019}.
The localization of FRB~180924 and FRB~190523 to massive galaxies could indicate that they differ in nature from FRB~121102, which is associated with a star-forming dwarf galaxy \citep{tendulkar2017}.
Since FRB~121102 and its environment is well studied, we go on to further discuss the implications of our work with regard to this source.

The host galaxy of FRB 121102 is a low-metallicity star-forming dwarf galaxy with a stellar mass of $\sim(4-7)\times10^7$\,M$_\odot$ \citep{tendulkar2017}. The FRB is coincident with a compact ($\lesssim 0.7$\,pc) persistent radio source \citep{marcote2017} which resides in a larger ($\approx$ 0.7\,kpc), bright star-formation region \citep{kokubo2017, bassa2017}. 

Superluminous supernovae and long GRBs are also found in SFRs of similar low-metallicity galaxies, strengthening the connection between sources, such as FRB 121102, and young neutron stars. The FRB's measured DM of 558\,pc\,cm$^{-3}$ can be explained by coming from the Milky Way, the IGM, and the host in roughly equal parts. Subtracting the expected DM$_{\rm MW}\approx220$\,pc\,cm$^{-3}$, and DM$_{\rm IGM}\approx115-285$\,pc\,cm$^{-3}$ from the total leaves $55-225$\,pc\,cm$^{-3}$ to be explained by the host. 
\refbf{
Applying the wind contribution as discussed in Sect. \ref{sec:pwn} to this excess DM, we infer that a neutron star powering FRB 121102 with a reasonably high magnetic field ($B\sim5\times10^{13}$\,G), spinning with $P~\sim100$\,ms, would generate a wind dispersion of $150$\,pc\,cm$^{-3}$ and can hence explain the host contribution.
}
However, this local excess DM could \refbf{also} come from the persistent radio source or from the SFR \citep{kokubo2017, bassa2017}. 
Using H$\alpha$ surface density of the SFR, \citet{bassa2017} argued that such a region could in principle contribute up to 589\,pc\,cm$^{-3}$. This may eschew the need for
significant DM contribution from the PWN or SNR, which would also explain the lack of strong secular changes in FRB 121102's observed DM. 

Although the FRB~121102 host galaxy is very different from our own, 
the environments of the associated pulsars in our Galactic sample could provide hints 
about the nature of the local dispersion in FRB~121102 and the high rotation measure (RM)  of $\sim 10^5$\,rad\,m$^{-2}$ that was measured \citep{michilli2018}.
Galactic pulsar J1550$-$5418, for example, is the youngest of our sample, 
and has the largest discrepancy between expected and measured DM found 
in our set of SNR pulsars with independent distance measurements. 
Its high DM of 830\,pc\,cm$^{-3}$, exceeding the prediction of 214\,pc\,cm$^{-3}$ three times, 
may in part be due to a local SFR \citep{gelfand2007}. 
One hint for this comes from its RM. It is one of the largest RMs of any known pulsar, at $-$1860\,rad\,m$^{-2}$. But that high RM is unlikely to be due to its local nebula, since the entire region within $\sim5$\,deg$^2$ of the pulsar also has extraordinarily large negative RM \citep{oppermann15}. 
Assuming that this same patch of free electrons -- a local SFR -- contributes to 
both to the large DM and the large RM indicates that in this case the nearby star-formation is the likely culprit.

\section{Summary}
\label{sec:conclusions}
In this work, we explore whether SNRs and PWNe contribute to the observed dispersion measure of the radio pulsars they host. % associated with SNRs and PWNe show a contribution to their dispersion measures from these hosts. 
We find an observed DM excess of $21.1 \pm 10.6$ pc\,cm$^{-3}$, which we conclude can be best explained by the doppler-boosted PWN ejecta combined with the ionized ejecta in SNRs. For these media we find an average contribution of $\sim 14$ pc\,cm$^{-3}$ for the sources in our sample.
No positive offset was observed in the comparison sample of pulsars not associated with such hosts. 
We further explored whether similar mechanisms can explain the DM excess observed in FRBs in the case when they arise from young pulsars.
Of course, there is no reason why all FRBs should be locally dispersed in the same way, given that they can come from different host galaxy types and have different environments.
The contribution of a wind nebula depends strongly on the parameters of that specific neutron star (see Eq.~\ref{eq-pwn}). The DM excess from a SNR depends on the source age; and for SFRs, different lines of sight may vary significantly in the number and nature of \ion{H}{ii} regions that they intersect.
That said, our study into local, galactic, associated pulsars finds a DM-excess trend that, 
when extrapolated to younger sources and more extreme environments, 
shows how FRBs can incur appreciable amounts of local  dispersion if 
  they arise from  young neutron stars. 
%Synthesizing these above aspects, we conclude that the doppler-boosted PWN ejecta combined with the ionized ejecta in SNRs provide the best explanation for the observed DM excess of $21.1 \pm 10.6$ pc\,cm$^{-3}$.  In radio pulsars associated with such hosts. No offset was observed in the comparison sample.

\begin{acknowledgements}
We would like to thank Joel M. Weisberg, Kohta Murase, Elena Amato, and Ralph Wijers for useful discussions. The research leading to these results has received funding from the European Research Council under the European Union’s Seventh
Framework Programme (FP/2007-2013)/ERC grant Agreement No. 617199 (ALERT) and from the Netherlands Research School for Astronomy (NOVA4-ARTS).
SMS was supported by the National Aeronautics and Space Administration (NASA) under grant number NNX17AL74G issued through the NNH16ZDA001N Astrophysics Data Analysis Program (ADAP).
We acknowledge the use of the python packages scipy, \citep{scipy}, matplotlib, \citep{matplotlib} and numpy \citep{numpy} throughout this work.
We would like to thank the anonymous referee whose thorough comments greatly improved this work.
\end{acknowledgements}

% for the bibliography, at the end 
\bibliographystyle{aa} % style aa.bst 
%\bibliographystyle{yahapj} % Makes hyperlinked refs
%\begin{thebibliography}{}
% \expandafter\ifx\csname natexlab\endcsname\relax\def\natexlab#1{#1}\fi
\bibliography{bibliography}

\end{document}